\documentclass[11pt,a4paper,USenglish]{article}

\providecommand{\keywords}[1]{\textbf{\textit{Keywords---}} #1}
\newcommand*{\qed}{\hfill\ensuremath{\square}}
\newenvironment{sketch}{\noindent{\em Sketch.~}}{\hspace*{\fill}${\small\qed}$\vspace{2mm}}
\newenvironment{proof}{\noindent{\em Proof.~}}{\hspace*{\fill}${\small\qed}$\vspace{2mm}}

\newif\ifshortversion

\shortversionfalse
\shortversiontrue

\ifshortversion
  \usepackage{environ}
  \NewEnviron{hide}{}

\else
\fi

\newcommand{\collapsed}{canonical digraph\xspace}
\newcommand{\canonical}{canonical c-graph\xspace}

\date{}
\usepackage{setspace}
\usepackage[margin=2.5cm]{geometry}
\usepackage{microtype}

\usepackage{tocloft}
\usepackage{ntheorem}
\theoremstyle{plain}
\theoremsymbol{}
\theoremprework{}
\theorempostwork{}
\theoremseparator{.}

\newtheorem{property}{Property}

\newtheorem{observation}{Observation}
\newtheorem{theorem}{Theorem}
\newtheorem{lemma}{Lemma}
\newtheorem{claim}{Claim}
\newtheorem{corollary}{Corollary}
\newtheorem{definition}{Definition}

\usepackage{subfig}
\usepackage{amsmath,amssymb}

\usepackage{dsfont}
\usepackage{xspace}
\usepackage[colorlinks=true]{hyperref}
\usepackage{paralist}
\usepackage{wrapfig}
\usepackage{floatrow}
\usepackage{rotating}
\usepackage{algorithm}
\usepackage[noend]{algpseudocode}
\usepackage{stmaryrd}
\usepackage{stackengine}
\usepackage{scalerel}

\newcommand{\etal}{{\em et al.}\xspace}

\renewcommand{\paragraph}[1]{\smallskip\noindent\textbf{#1}\xspace}
\graphicspath{{figures/}}

\usepackage[nameinlink,capitalise]{cleveref}
\Crefname{observation}{Observation}{Observations}
\Crefname{algorithm}{Algorithm}{Algorithms}
\Crefname{section}{Section}{Sections}
\Crefname{observation}{Observation}{Observations}
\Crefname{lemma}{Lemma}{Lemmas}
\Crefname{claim}{Claim}{Claims}
\Crefname{figure}{Fig.}{Figs.}
\Crefname{figure}{Fig.}{Figs.}
\Crefname{enumi}{Condition}{Conditions}
\Crefname{property}{Property}{Properties}

\newcommand{\appendixIrreducible}{\textcolor{red}{Section~9 [full version]}\xspace}
\newcommand{\appendixHardness}{\textcolor{red}{Section~10 [full version]}\xspace}
\newcommand{\appendixOuterplanar}{\textcolor{red}{Section~11 [full version]}\xspace}
\newcommand{\appendixNAESAT}{\textcolor{red}{Section~8 [full version]}\xspace}

\usepackage{color}
\definecolor{realblue}{rgb}{0,0,1}
\definecolor{blue}{rgb}{0.274,0.392,0.666}
\definecolor{darkerblue}{rgb}{0.094,0.455,0.804}
\definecolor{darkblue}{rgb}{0.063,0.306,0.545}
\definecolor{red}{rgb}{0.627,0.117,0.156}
\definecolor{green}{rgb}{0,0.588,0.509}
\definecolor{orange}{rgb}{0.903,0.739,0.382}
\definecolor{realred}{rgb}{1,0,0}

\newcommand{\red}[1]{{{\textcolor{red}{#1}\xspace}}}

\newcommand{\blue}[1]{{{\textcolor{darkerblue}{#1}\xspace}}}

\renewcommand{\emph}[1]{\blue{\em #1}}

\hypersetup{linkcolor={red},citecolor={red},urlcolor={red}} 

\newcommand{\calT}[1]{\ensuremath{{\cal T}^{#1}}\xspace}

\newcommand{\NAESAT}{{\sc NAESAT}\xspace}

\newcommand{\NAE}{{\sc NAE}\xspace}

\newcommand{\inn}{
\scaleobj{0.7}{
\ensurestackMath{\stackon[-.25mm]{\shortuparrow}{\scaleobj{0.9}{\circ}}}}\xspace
}

\newcommand{\out}{
\begin{turn}{180}
\scaleobj{-0.7}{
{\ensurestackMath{\stackon[0pt]{\scaleobj{0.9}{\circ}}{\shortuparrow}}}
}
\end{turn}\xspace
}

\renewcommand{\downarrow}{\inn}
\renewcommand{\uparrow}{\out}

\DeclareMathOperator{\pert}{pert}
\DeclareMathOperator{\skel}{skel}

\newcommand{\even}[1]{\ensuremath{\mathds{E}^+_{#1}}}

\newcommand{\maxmodality}{\mbox{\sc $4$-MaxModality}\xspace}
\newcommand{\modality}{\mbox{\sc $4$-Modality}\xspace}

\newcommand{\maxkmodality}[1]{\mbox{\sc $#1$-MaxModality}\xspace}
\newcommand{\kmodality}[1]{\mbox{\sc $#1$-Modality}\xspace}

\newcommand{\problemdef}[3]{
\medskip
\noindent
	\begin{tabular}{|p{.968\textwidth}|}
	\hline
	{\bf \noindent  \textbf{Problem:} {\sc #1}}
	\\
	\hline
	\hline
	\vspace{-6mm}
	\begin{description}
		\item[Input:] #2
		\vspace{-2mm}
		\item[Question:] #3
		\vspace{-4mm}
	\end{description}
	\\
	\hline
	\end{tabular}
}

\begin{document}
\title{Computing $k$-Modal Embeddings of Planar Digraphs}

\author{%
  {Juan Jos\'e~Besa}%
    \thanks{Computer Science Department, University of California, Irvine, California, USA \protect\url{jjbesavi@uci.edu}.}
\and
  {Giordano {Da Lozzo}}%
    \thanks{Department of Engineering, Roma Tre University, Rome, Italy \protect\url{giordano.dalozzo@uniroma3.it}.}
\and
  {Michael T. Goodrich}%
    \thanks{Computer Science Department, University of California, Irvine, California, USA \protect\url{goodrich@uci.edu}.}}

\maketitle
\thispagestyle{empty}
\begin{abstract}
Given a planar digraph $G$ and a positive even integer $k$, an embedding of $G$ in the plane is \emph{$k$-modal}, if every vertex of $G$ is incident to at most $k$ pairs of consecutive edges with opposite orientations, i.e., the incoming and the outgoing edges at each vertex are grouped by the embedding into at most $k$ sets of consecutive edges with the same orientation. 
In this paper, we study the \kmodality{k} problem, which asks for the existence of a $k$-modal embedding of a planar digraph. 
This combinatorial problem is at the very core of a variety of constrained embedding questions for planar digraphs and flat clustered networks.

First, since the \kmodality{2} problem can be easily solved in linear time,
we consider the general \kmodality{k} problem for any value of $k>2$ and show that
the problem is NP-complete for planar digraphs of maximum degree $\Delta \geq k+3$. We relate its computational complexity to that of two notions of planarity for flat clustered networks: Planar Intersection-Link and Planar NodeTrix representations. 
This allows us to answer in the strongest possible way an open question by \mbox{Di Giacomo} \etal~\href{https://doi.org/10.1007/978-3-319-73915-1_37}{[GD17]}, concerning the complexity of constructing planar NodeTrix representations of flat clustered networks with small clusters, and to address a
research question by Angelini \etal~\href{https://doi.org/10.7155/jgaa.00437}{[JGAA17]}, concerning intersection-link representations based on geometric objects that determine complex arrangements.
On the positive side,
we provide a simple FPT algorithm for partial $2$-trees of arbitrary degree, whose running time is exponential in $k$ and linear in the input size.  
 
Second, motivated by the recently-introduced planar \mbox{L-drawings} of planar digraphs~\href{https://doi.org/10.1007/978-3-319-73915-1_36}{[GD17]}, which require the computation of a $4$-modal embedding, we 
focus our attention on $k=4$. On the algorithmic side, we show a complexity \mbox{dichotomy} for the \modality problem with respect to~$\Delta$, by providing a linear-time algorithm for planar digraphs with~\mbox{$\Delta \leq 6$}. 
This algorithmic result is based on
decomposing the input digraph into its blocks via BC-trees and each of these blocks into its triconnected components via SPQR-trees. In particular, we are able to show that the constraints imposed on the embedding by the rigid triconnected components can be tackled by means of a small set of reduction rules and discover that the algorithmic core of the problem lies in special instances of \mbox{\NAESAT}, which we prove to be always \NAE-satisfiable---a result \mbox{of independent interest that improves on Porschen \etal~\href{https://doi.org/10.1007/978-3-540-24605-3_14}{[SAT03]}.}

Finally, on the combinatorial side, we consider outerplanar digraphs and show that any such a digraph always admits a $k$-modal embedding with $k=4$ and that this value of $k$ is best possible for the digraphs in this family.
\end{abstract}



\clearpage
\section{Introduction}\label{se:intro} 
\begin{figure}[b!]
\centering
\subfloat[\label{fig:l-drawing-example}]{\includegraphics[page=1,height=0.2\textwidth]{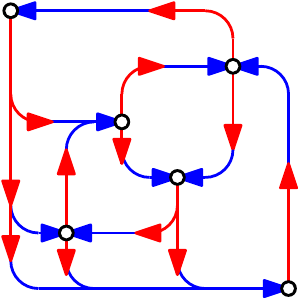}}\hfil
\subfloat[\label{fig:nodetrix-example}]{\includegraphics[height=0.2\textwidth,page=1]{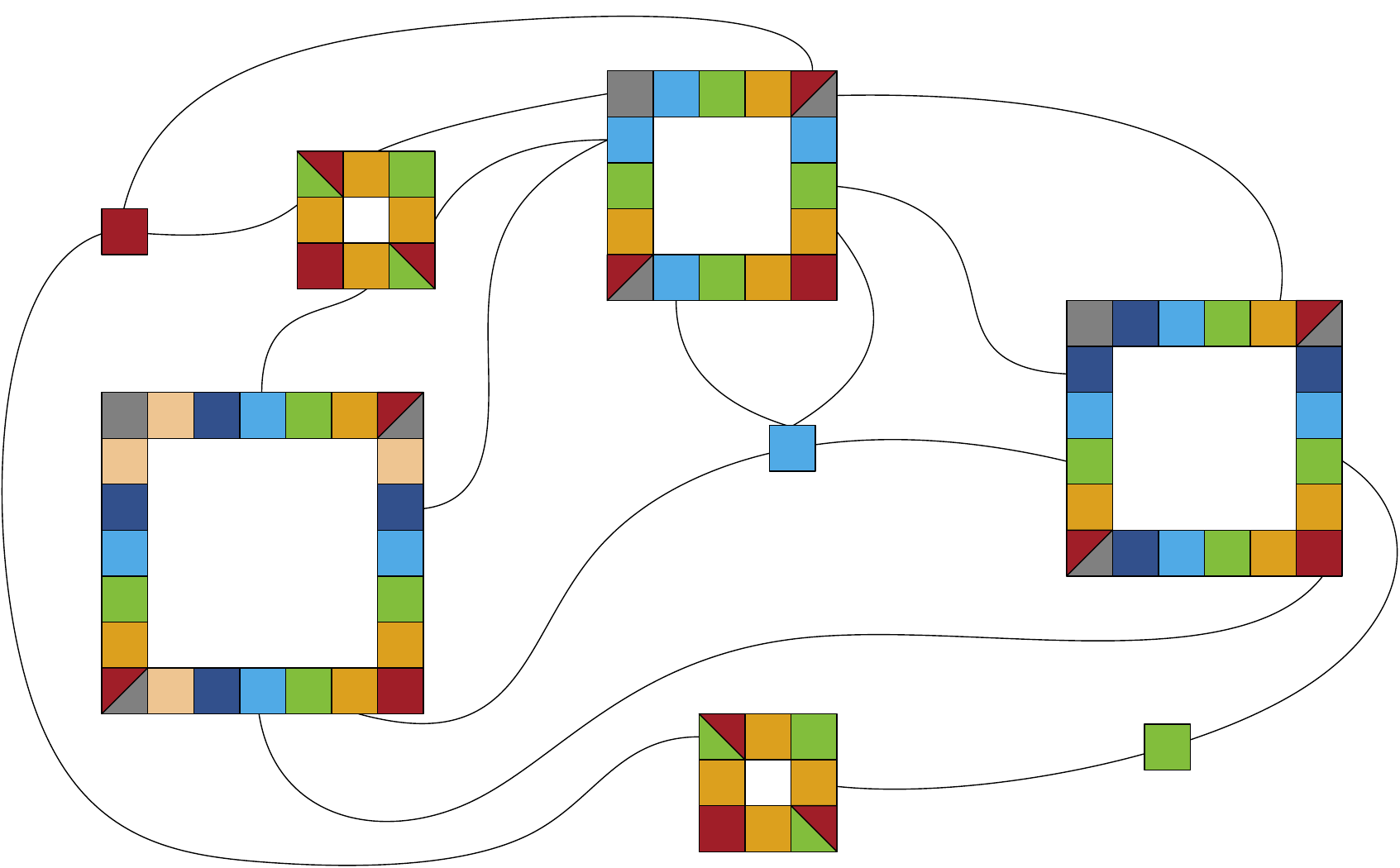}}\hfil
\subfloat[\label{fig:clique-example}]{\includegraphics[page=1,height=0.2\textwidth]{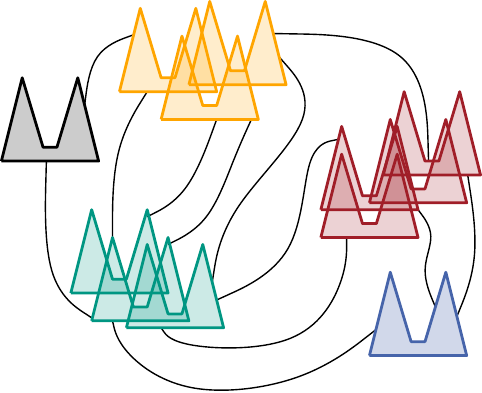}}
\caption{
(a) A planar L-drawing, which determines a $4$-modal embedding. (b) A planar NodeTrix representation. (c) A planar intersection-link representation using comb-shaped polygons.
}
\end{figure}
%
Computing $k$-modal embeddings of planar digraphs, for some positive even integer $k$ called \emph{modality}, is an important algorithmic task at the basis of several types of graph visualizations. 
In \emph{$2$-modal embeddings}, also called \emph{bimodal embeddings}, the outgoing and the incoming edges at each vertex form two disjoint sequences. 
Bimodal embeddings are ubiquitous in Graph Drawing. 
For instance, level planar drawings~\cite{BattistaN88,JungerLM98} and upward-planar drawings~\cite{BattistaT87,GargTamassia01}---two of the most deeply-studied graph drawing standards---determine bimodal embeddings. 
\emph{4-modal embeddings}, where the outgoing and the incoming edges at each vertex form up to four disjoint sequences with alternating orientations, arise in the context of planar \mbox{L-drawings} of digraphs. In an \emph{L-drawing} of an $n$-vertex digraph, introduced by Angelini \etal~\cite{AngeliniLBDPRT16}, vertices are placed on the $n \times n$~grid so that each vertex is assigned a unique $x$-coordinate and a unique $y$-coordinate and each edge $uv$ (directed from $u$ to $v$) is represented as a $1$-bend orthogonal polyline composed of a vertical segment incident to $u$ and of a horizontal segment incident to $v$.  
Recently, Chaplick \etal~\cite{ccdnptw-plddg-17} addressed the question of deciding the existence of \emph{planar L-drawings}, i.e., L-drawings whose edges might possibly overlap but do not cross and observe that the existence of a $4$-modal embedding is a necessary condition for a digraph to admit such a representation (\cref{fig:l-drawing-example}).

To the best of our knowledge, no further relationships have been explicitly pointed out in the literature between modal embeddings and notable drawing models for modality values greater than four, yet they do exist.
\mbox{Da Lozzo} \etal~\cite{LozzoBFP18} and Di Giacomo \etal~\cite{GiacomoLPT17} study the planarity of \emph{NodeTrix representations} of flat clustered networks, a hybrid representational model introduced by Henry, Fekete, and McGuffin~\cite{DBLP:journals/tvcg/HenryFM07}, where clusters and intra-cluster edges are represented as adjacency-matrices, with rows and columns for the vertices of each cluster, and inter-cluster edges are Jordan arcs connecting different matrices (\cref{fig:nodetrix-example}).
For clusters containing only two vertices, it is possible to show that the problem of computing planar NodeTrix representations coincides with the one of testing whether a special digraph, called the \emph{\collapsed}, associated to the network admits a $6$-modal embedding. 
For higher values of modality, $k$-modal embeddings occur in the context of Intersection-Link representations of flat clustered networks. In an \emph{intersection-link representation}~\cite{DBLP:journals/jgaa/AngeliniLBFPR17,DBLP:conf/gd/AngeliniEH0KLNT18}, vertices are represented as translates of the same polygon, intra-cluster edges are represented via intersections between the 
polygons corresponding to their endpoints, and inter-cluster edges---similarly to NodeTrix representations---are Jordan arcs connecting the polygons corresponding to their endpoints. For any modality $k\geq 2$, it can be shown that testing the existence of a $k$-modal embedding of the \collapsed of a flat clustered network with clusters of size two is equivalent to testing the existence of an intersection-link representation in which the curves representing inter-cluster edges do not intersect, when vertices are drawn as comb-shaped polygons (\cref{fig:clique-example}).

{\bf Related Work.}
It is common knowledge that the existence of bimodal embeddings can be tested in linear time: Split each vertex $v$ that has both incoming and outgoing edges into two vertices $v_{in}$ and $v_{out}$, assign the incoming edges~to~$v_{in}$ and the outgoing edges to $v_{out}$,
connect $v_{in}$ and $v_{out}$ with an edge, and test the resulting (undirected) graph \mbox{for planarity using any of} the linear-time planarity-testing algorithms~\cite{BoothL76,HopcroftT74}. 
Despite this, most of the planarity variants requiring bimodality are NP-complete; for instance, upward planarity~\cite{GargTamassia01}, windrose planarity~\cite{AngeliniLBDKRR16}, {partial-level planarity}~\cite{BrucknerR17},
clustered-level planarity and $T$-level planarity~\cite{Angelini15,KlemzR17}, ordered-level planarity and bi-monotonicity~\cite{KlemzR17}. In this scenario, a notable exception is represented by the classic level planarity problem, which can be solved in linear time~\cite{JungerLM98}, and its generalizations on the standing cylinder~\cite{BachmaierBF05}, rolling cylinder and the torus~\cite{AngeliniLBFPR16}. 
Although the existence of a bimodal embedding is easy to test,  Binucci, Didimo, and Giordano~\cite{BinucciDG08} prove that the related problem of finding the maximum bimodal subgraph of an embedded planar digraph is an NP-hard problem.
Moreover, Binucci, Didimo, and Patrignani~\cite{BinucciDP14} show that, given a \emph{mixed
planar graph}, i.e., a planar graph whose edge set is partitioned into a set of directed edges and a set of undirected edges, orienting the undirected edges in such a way that the whole graph admits a bimodal embedding is an NP-complete problem. 
On the other hand, the question regarding the computational complexity of constructing $k$-modal embedding for $k\geq 4$ has not been addressed, although the related problem of testing the existence of \mbox{planar L-drawings has been recently proved NP-complete~\cite{ccdnptw-plddg-17}.}

\paragraph{\bf Our results.}
We study the complexity of the \kmodality{k} problem, which asks for the existence of $k$-modal embeddings of planar digraphs---with an emphasis on $k=4$. \mbox{Our results are as follows:}
\begin{enumerate}[-]
	\item We demonstrate a complexity dichotomy for the \modality problem with respect to the maximum degree $\Delta$ of the input digraph. Namely, we show NP-completeness when $\Delta \geq 7$
	\ifshortversion{{\mbox(see \appendixHardness)} }\else{(\cref{th:npc}) }\fi 
	and give a linear-time testing algorithm for $\Delta \leq 6$ (\cref{th:max-degree}). Further, we extend the hardness result to any modality value larger than or equal to $4$, by proving that the \kmodality{k} problem is NP-complete for $k\geq 4$ when $\Delta \geq k+3$.
	\item We provide an FPT-algorithm for \kmodality{k} that runs in $f(k)O(n)$ time for the class of directed partial $2$-trees (\cref{th:2trees-efficient}), which includes series-parallel and outerplanar digraphs. 
	\item In \cref{se:relationship}, we relate $k$-modal embeddings with hybrid representations of flat clustered graphs, and exploit this connection to give new complexity results (\cref{th:nodetrix-hard,th:clique-hard}) and algorithms (\cref{th:nodetrix-polynomial,th:clique-planarity-polynomial}) for these types of representations. In particular, our NP-hardness results allow us to
	answer two open questions. Namely, we settle in the strongest possible way an open question, posed by Di Giacomo \etal~\cite[Open~Problem~(i)]{GiacomoLPT17}, about the complexity of computing planar NodeTrix representations of flat clustered graphs with clusters of size smaller than $5$. Also, we address a research question by Angelini \etal~\cite[Open~Problem~(2)]{DBLP:journals/jgaa/AngeliniLBFPR17} about the representational power of intersection-link representations based on geometric objects that give rise to complex combinatorial structures, and solve it when the considered geometric objects are $k$-combs.
	\item Finally, 
	in \ifshortversion{\appendixOuterplanar}\else{\cref{apx:outerplanar}}\fi, we show that not every outerplanar digraph admits a bimodal embedding, whereas any outerplanar (multi-)digraph admits a $4$-modal embedding.
\end{enumerate}
The algorithms presented in this paper employ the SPQ- and SPQR-tree data structures to succinctly represent the exponentially-many embeddings of series-parallel and biconnected planar digraphs, respectively, and can be easily modified to output an embedding of the input digraph in the same time bound. 
In particular, our positive result for $\Delta \leq 6$ is based on a set of simple reduction rules that exploit the structure of the rigid components of bounded-degree planar digraphs. These rules allow us to tackle the algorithmic core of the problem, by enabling a final reduction step to special instances of \NAESAT, previously studied by Porschen \etal~\cite{PorschenRS03},
which we prove to be always \NAE-satisfiable \ifshortversion{\appendixNAESAT}\else{in \cref{sse:naesat}\fi.\footnote{In ``{\em Stefan Porschen, Bert Randerath, Ewald Speckenmeyer: Linear Time Algorithms for Some Not-All-Equal Satisfiability Problems. SAT 2003: 172-187}''~\cite{PorschenRS03}, the authors state in the abstract ``First we show that a \NAESAT model (if existing) can be computed in linear time for formulas in which each variable occurs at most twice.''. We give a strengthening of this result by showing that the only negative formulas with the above properties are those whose variable-clause graph contains components isomorphic to a simple cycle and provide a recursive linear-time algorithm for computing a \NAE-truth assignment for formulas in which each variable occurs at most twice, when one exists, which is also considerably simpler than the one presented in~\cite{PorschenRS03}.}


\section{Definitions}

We assume familiarity with basic concepts concerning directed graphs, planar embeddings, connectivity and the BC-tree data structure; see the~full~version for more details.

\paragraph{Directed graphs.} 
A \emph{directed graph} (for short \emph{digraph}) $G=(V,E)$ is a pair, where $V$ is the set of vertices and $E$ is the set of \emph{directed edges} of~$G$, i.e., ordered pairs of vertices. We also denote sets $V$ and $E$ by $V(G)$ and $E(G)$, respectively.
The \emph{underlying graph} of $G$ is the undirected graph obtained from $G$ by disregarding edge directions.
Let $v$ be a vertex, we denote by $E(v)$ the set of edges of $G$ incident to $v$ and by $\deg(v)=|E(v)|$ the \emph{degree} of $v$.
For an edge $e=uv$ and an end-point $x \in \{u,v\}$ of $e$, we define the \emph{orientation} $\sigma(e, x)$ of $e$ \emph{at} $x$ as $\sigma(e, x)= \out$, if $x=u$, and $\sigma(e, x)= \inn$, if $x=v$, and we say that $uv$ is \emph{outgoing from $u$} and \emph{incoming at $v$}. 

\paragraph{\bf Modality.}%
Let $G$ be a planar digraph and let $\cal E$ be an embedding of $G$.
A pair of edges $e_1,e_2$ that appear consecutively in the circular order around a vertex $v$ of $G$ is \emph{alternating} if they do not have the same orientation at $v$, i.e., they are not both incoming at or both outgoing from $v$. Also, we say that vertex $v$ is \emph{$k$-modal}, or that $v$ has \emph{modality $k$}, or that the \emph{modality of} $v$ \emph{is $k$} in~$\cal E$,  if there exist exactly $k$ alternating pairs of edges incident to $v$ in $\cal E$. 
Clearly, the value $k$ needs to be a non-negative even integer.
An embedding of a digraph $G$ is \emph{$k$-modal}, if each vertex is at most $k$-modal; see \red{ \cref{fi:spqr}(left)}.

We now define an auxiliary problem, called {\sc $k$-MaxModality} (where $k$ is a positive even integer), which will be useful to prove our algorithmic results. We denote the set of non-negative integers by $\mathds{Z}^*$ and the set of non-negative even integers smaller than or equal to $k$ as~$\even{k} = \{b: b=2a, b \leq k, a \in \mathds{Z}^* \}$.
Given a graph $G$, we call \emph{maximum-modality function} an integer-valued function $m: V(G) \rightarrow \even{k}$.  We say that an embedding $\cal E$ of $G$ \emph{satisfies} $m$ \emph{at a vertex} $v$ if the modality of $v$ in $\cal E$ is at most~$m(v)$.

\problemdef{\sc $k$-MaxModality}{A pair $\langle G, m\rangle$, where $G$ is a digraph and $m$ is a maximum-modality function.}{Is there an embedding $\cal E$ of $G$ that \emph{satisfies} $m$ at every vertex?}

\section{Implications on Hybrid Representations}\label{se:relationship}

A \emph{flat clustered graph} (for short, \emph{c-graph}) is a pair $\mathcal C = (G=(V,E),\mathcal P=(V_1,V_2,\dots,V_c))$, where $G$ is a graph and $\mathcal P$ is a partition of $V$ into sets $V_i$, for $i=1,\dots,c$, called \emph{clusters}\ifshortversion{. }\else{\footnote{The more general notion of \emph{clustered graph} is obtained by allowing the set of clusters to form a laminar set family, which is better described by a rooted tree $\mathcal T$ 
whose leaves are the vertices of $G$ and whose every internal node $\mu$ represents the cluster containing the leaves of the subtree of $\mathcal T$ rooted at $\mu$. However, since a flat clustering is more naturally described by a partition, rather then by a tree, we define c-graphs using $\mathcal P$ rather then $\mathcal T$.}. }\fi An edge $(u, v) \in E$ with $u \in V_i$ and $v \in V_j$ is an \emph{intra-cluster edge}, if $i = j$, and is an \emph{inter-cluster edge}, if $i \neq j$.
\begin{figure}[tb!]
\centering
\subfloat[\label{fi:geometric-correspondance}]{\includegraphics[page=3,height=0.21\textwidth]{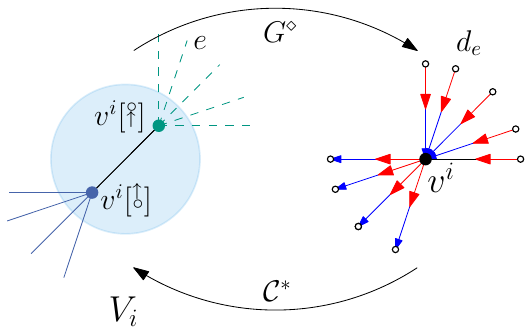}}\hfill
\subfloat[\label{fi:6-modal-node}]{\includegraphics[page=4,height=0.21\textwidth]{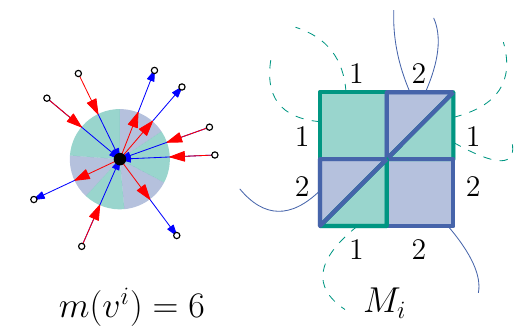}}\hfill
\subfloat[\label{fi:4-modal-intersection}]{\includegraphics[page=5,height=0.21\textwidth]{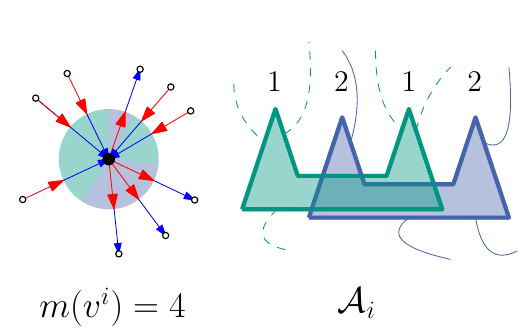}}
\caption{
(a) Illustrations for the duality between the \collapsed and the \canonical.
Correspondence (b) between $6$-modal embeddings and planar NodeTrix representations, and (c) between
$4$-modal embeddings and clique-planar representations \mbox{using $2$-combs as geometric objects}.}
\end{figure}
The problem of visualizing such graphs so to effectively convey both the relation information encoded in the set $E$ of edges of $G$ and the hierarchical information given by the partition $\mathcal P$ of the clusters has attracted considerable research attention.
\ifshortversion{}\else{As crossing-free graph drawings are universally considered more readable~\cite{DBLP:conf/gd/Purchase97,DBLP:journals/iwc/Purchase00}, 
this effort has culminated in several notions of planarity for c-graphs. The most celebrated of such notions, introduced by Feng, Cohen, and Eades~\cite{DBLP:conf/esa/FengCE95}, goes by the name of {\sc Clustered Planarity}
and asks for the existence of a \emph{c-planar drawing} of a c-graph, that is, a planar drawing of $G$ together with a representation of each cluster $V_i$ as a region of the plane homeomorphic to a closed disk that contains the drawing of the subgraph of $G$ induced by cluster $V_i$; additionally, clusters may not intersect each other and edges may cross the boundary of each cluster at most once. Alongside the classical notion of c-planar drawings, new hybrid models for the visualization of flat clustered networks (and corresponding planarity notions) have recently received considerable attention. 
}\fi
In a \emph{hybrid representation} of a graph different conventions are used to represent the dense and the sparse portions of the graph~\cite{DBLP:journals/jgaa/AngeliniLBFPR17,DBLP:conf/gd/AngeliniEH0KLNT18,LozzoBFP18,GiacomoLPT17,DBLP:journals/tvcg/HenryFM07,DBLP:conf/IEEEscc/LuZXXL14,DBLP:journals/tvcg/YangSDTLT17}. We present important implications of our results on some well-known \mbox{models for hybrid-representations of c-graphs.}

Let $\mathcal C$ be a c-graph whose every cluster forms a clique of size at most $2$, that is, each cluster contains at most two vertices connected by an intra-cluster edge. 
Starting from $\mathcal C$ we define an auxiliary digraph $G^\diamond$, called the \emph{\collapsed} for $\mathcal C$, as follows. Without loss of generality, assume that,
for $i=1,\dots,c$, each cluster $V_i$ contains two vertices denoted as $v^i[\inn]$ and $v^i[\out]$.
The vertex set of $G^\diamond$ contains a vertex $v^i$, for $i=1,2,\dots,c$, and a dummy vertex $d_e$, for each inter-cluster edge $e \in E$.
The edge set of $G^\diamond$ contains two directed edges, for each inter-cluster edge $e = (v^i_x,v^j_y) \in E$, with $x,y \in \{\inn,\out\}$ and $i \neq j$; namely, $E(G^\diamond)$ contains (i) 
either the directed edges $v^i_x d_e$, if $x = \out$, or the directed edge $d_e v^i_x$, if $x = \inn$, and 
(ii) either the directed \mbox{edges $v^i_y d_e$, if $y = \out$, or the directed edge $d_e v^i_y$, if $y = \inn$.}

Let now $D=(V,E)$ be a digraph. We construct a c-graph $\mathcal C^*=(G^*=(V^*,E^*),\mathcal P^*)$ from $D$ whose every cluster forms a clique of size at most $2$, called the \emph{\canonical} for $D$, as follows. For each vertex $v^i \in V$, $G^*$ contains two vertices $v^i[\inn]$ and $v^i[\out]$, which form the cluster $V_i = \{v[\inn],v[\out]\}$ in $\mathcal P^*$.
For each (directed) edge $v^i v^j$ of $D$,  $G^*$ contains an (undirected) edge $(v^i[\out],v^j[\inn])$; that is, each directed edge in $E$ that is incoming (outgoing) at a vertex $v^i$ and outgoing (incoming) at a vertex $v^j$ corresponds to an inter-cluster edge in $E^*$ incident to $v^i[\inn]$ (to $v^i[\out]$) and to $v^j[\out]$ (to $v^j[\inn]$).
Finally, for each vertex $v^i \in V$, $G^*$ contains an intra-cluster edge $(v^i[\inn],v^i[\out])$. 
The \collapsed and the \canonical form dual concepts, as illustrated in \cref{fi:geometric-correspondance}; the \canonical of $G^\diamond$ is the original c-graph $\cal C$ (neglecting clusters originated by dummy vertices) and the \collapsed of $\mathcal C^*$ is the original \mbox{digraph $D$ (suppressing dummy vertices).}

\smallskip
\paragraph{NodeTrix Planarity.} 
A \emph{NodeTrix representation} of a c-graph $\mathcal C = (G,\mathcal P)$ is a drawing of $\mathcal C$ such that: 
\begin{inparaenum}[\bf (i)]
\item Each cluster $V_i \in \mathcal P$ is represented as a symmetric adjacency matrix $M_i$ (with $|V_i|$ rows and columns), drawn in the plane so that its boundary is a square $Q_i$ with sides parallel to the coordinate axes. 
\item No two matrices intersect, that is, $Q_i \cap Q_j = \emptyset$, for all $1 \leq i < j \leq c$.
\item Each intra-cluster edge is represented by the adjacency matrix $M_i$. 
\item Each inter-cluster edge $(u,v)$ with $u \in V_i$ and $v \in V_j$ is represented as a simple Jordan arc connecting a point on the boundary of $Q_i$ with a point on the boundary of $Q_j$, where the point on $Q_i$ (on $Q_j$) belongs to the column or to the row of $M_i$ (resp. of $M_j$) associated with $u$ (resp. with $v$). 
A NodeTrix representation is \emph{planar} if no inter-cluster edge intersects a matrix or another inter-cluster edge, except possibly at a common end-point; see \cref{fig:nodetrix-example,fi:6-modal-node}. 
The {\sc NodeTrix Planarity} problem asks whether a c-graph admits a planar NodeTrix representation.
{\sc NodeTrix Planarity} has been proved NP-complete for c-graphs whose clusters have size larger than or equal to $5$~\cite{GiacomoLPT17}.

\end{inparaenum}

We are ready to establish our main technical lemmas.

\begin{lemma}\label{lem:nodetrix-6-modal}
C-graph $\mathcal C$ is planar NodeTrix if and only if $G^\diamond$ admits a $6$-modal embedding.
\end{lemma}

\begin{lemma}\label{lem:6-modal-nodetrix}
Digraph $D$ admits a $6$-modal embedding if and only if $\mathcal C^*$ is planar NodeTrix.
\end{lemma}

\paragraph{Proof sketch for \cref{lem:nodetrix-6-modal,lem:6-modal-nodetrix}.}
Let $M_i$ be the matrix representing cluster $V_i = \{v^i[\inn],v^i[\out]\}$. We have that, independently of which of the two possible permutations for the rows and columns of $M_i$ is selected, the boundary of $Q_i$ is partitioned into three maximal portions associated with $v^i[\inn]$ and three maximal portions associated with $v^i[\out]$; that is, they form the pattern $[1,2,1,2,1,2]$, see \cref{fi:6-modal-node}. Therefore, any planar NodeTrix representation of $\cal C$ (of $\cal C^*$) can be turned into a $6$-modal embedding of $G^\diamond$ (of $D$) via a local redrawing procedure which operates in the interior of $Q_i$; also, any $6$-modal embedding of $G^\diamond$ (of $D$) can be turned into a planar NodeTrix representation of $\cal C$ ($\cal C^*$) via a local redrawing procedure which operates in a small disk centered at $v_i$ that contains only $v_i$ and intersects only edges incident to $v_i$.

\smallskip 
Since $G^\diamond$ can be constructed in linear time from $\mathcal C$, \cref{lem:nodetrix-6-modal} and the algorithm of \cref{th:2trees-efficient} for solving \kmodality{k} of directed partial $2$-trees give us the following.

\begin{theorem}\label{th:nodetrix-polynomial}
{\sc NodeTrix Planarity} can be solved in linear time for flat clustered graphs whose clusters have size at most $2$ and whose \collapsed is a directed partial $2$-tree.
\end{theorem}

Note that 
(i) $\mathcal C^*$ can be constructed in polynomial time from $D$, 
(ii) $\mathcal C^*$ only contains clusters of size $2$ (although clusters corresponding to vertices of $D$ incident to incoming or outgoing edges only could be simplified into clusters of size $1$), and 
(iii) each cluster 
$V_i \in \mathcal P^*$, with $v^i \in V(D)$,
is incident to $\alpha$ inter clusters edges, where $\alpha$ is the degree of $v^i$ in $D$. 
These properties and the fact that in \ifshortversion{\appendixHardness}\else{\cref{th:npc}}\fi\xspace we prove the \kmodality{k} problem to be NP-complete for digraphs of maximum degree $\Delta \geq k + 3$ give us the following.

\begin{theorem}\label{th:nodetrix-hard}
{\sc NodeTrix Planarity} is NP-complete for flat clustered graphs whose clusters have size at most $2$, even if each cluster is incident to at most $9$ inter-cluster edges.
\end{theorem}

We remark that the above NP-completeness result is best possible in terms of the size of clusters, as clusters of size $1$ do not offer any advantage to avoid intersections between inter-cluster edges. Also, it solves \cite[Open~Problem~(i)]{GiacomoLPT17}, which asks for the complexity of {\sc NodeTrix Planarity} for c-graphs whose clusters have size between $2$ and $5$.

\smallskip
\paragraph{Clique Planarity.}
Hybrid representations have also been recently studied in the setting in which clusters are represented via intersections of geometric objects. In particular, Angelini \etal~\cite{DBLP:journals/jgaa/AngeliniLBFPR17} introduced the following type of representations. Suppose that a c-graph $(G,\mathcal P)$ is given, where $\mathcal P$ is {\em a set of cliques} that partition the vertex set of $G$. In an \emph{intersection-link representation}, the vertices of $G$ are represented by geometric objects that are translates of the same rectangle. 
Consider an edge $(u,v)$ and let $R(u)$ and $R(v)$ be the rectangles representing $u$ and $v$, respectively. 
If $(u,v)$ is an intra-cluster edge (called \emph{intersection-edge} in \cite{DBLP:journals/jgaa/AngeliniLBFPR17}), we represent it by drawing $R(u)$ and $R(v)$ so that they intersect, otherwise if $(u,v)$ is an intra-cluster edge
(called \emph{link-edge} in \cite{DBLP:journals/jgaa/AngeliniLBFPR17}), we represent it by a Jordan arc connecting $R(u)$ and $R(v)$. 
A \emph{clique-planar} representation is an intersection-link representation in which no inter-cluster edge intersects the interior of any rectangle or another 
inter-cluster edge, except possibly at a common end-point.
The {\sc Clique Planarity} problem asks whether a c-graph $(G, \mathcal P)$ admits a clique-planar representation.

Angelini \etal proved the {\sc Clique Planarity} problem to be NP-complete, when $\mathcal P$ contains a cluster $V^*$ with $|V^*| \in O(|G|)$, and asked, in \cite[Open~Problem~(2)]{DBLP:journals/jgaa/AngeliniLBFPR17}, 
about the implications of using different geometric objects for representing vertices, rather than translates of the same rectangle. 
We address this question by considering $k$-combs as geometric objects, where a \emph{$k$-comb} is the simple polygon with $k$ spikes illustrated in \cref{fi:4-modal-intersection}.
We have the following.

\begin{lemma}\label{lem:embedding-clique-planar}
C-graph $\mathcal C$ is a positive instance of {\sc Clique Planarity} using $k$-combs as geometric objects if and only if $G^\diamond$ admits a $2k$-modal embedding.
\end{lemma}

\begin{lemma}\label{lem:clique-planarity-k-modal}
Digraph $D$ admits an $4$-modal embedding if and only if
$\mathcal C^*$ is a positive instance of {\sc Clique Planarity} using $2$-combs as geometric objects.
\end{lemma}

\paragraph{Proof sketch for \cref{lem:embedding-clique-planar,lem:clique-planarity-k-modal}.}
Let $A_i$ be an arrangements of $2$-combs representing cluster $V_i = \{v^i[\inn],v^i[\out]\}$. We have that, the boundary of $A_i$ is partitioned into at most two maximal portions associated with $v^i[\inn]$ and at most two maximal portions associated with $v^i[\out]$; that is, they form the pattern $[1,2,1,2]$, see \cref{fi:4-modal-intersection}. Therefore, as for \cref{lem:nodetrix-6-modal,lem:6-modal-nodetrix}, we can exploit a local redrawing procedure to transform a clique-planar representation of $\cal C$ (of $\mathcal C^*$) into a $4$-modal embedding of $G^\diamond$ (of $D$), and vice versa.

\smallskip 
Combining \cref{lem:embedding-clique-planar} and the algorithm of \cref{th:2trees-efficient} gives us the following positive result.

\begin{theorem}\label{th:clique-planarity-polynomial}
{\sc Clique Planarity} using $r$-combs, with $r\geq 1$, as geometric objects  
can be solved in linear time for flat clustered graphs whose clusters have size at most $2$ and whose \collapsed is a directed partial $2$-tree.
\end{theorem}

\smallskip
Finally, \cref{lem:clique-planarity-k-modal} and the discussion preceding \cref{th:nodetrix-hard} imply the following.

\begin{theorem}\label{th:clique-hard}
{\sc Clique Planarity} using $2$-combs as geometric objects is NP-complete, even for flat clustered graphs with clusters of size at most $2$ each incident to at most $7$ inter-cluster edges.
\end{theorem}

\section{Polynomial-time Algorithms}\label{se:ptime}

In this section, we present an algorithmic framework to devise efficient algorithms for the \kmodality{k} problem for notable families of instances. First, 
in \cref{se:simply}, we show how to efficiently reduce the \kmodality{k} problem in simply-connected digraphs to the \maxkmodality{k} problem in biconnected digraphs. Then, in \cref{se:bico}, we introduce preliminaries and definitions concerning SPQR-trees and $k$-modal embeddings of biconnected digraphs. 

\subsection{Simply-Connected Graphs}\label{se:simply}
We first observe that the \maxkmodality{k} problem is a generalization of the \kmodality{k}problem. In fact, a directed graph $G=(V,E)$ admits a $k$-modal embedding if and only if the pair $\langle G, m \rangle$, with $m(v)=k, \forall v \in V(G)$, is a positive instance of the \maxkmodality{k} problem. 

\begin{observation}\label{obs:reduction}
\kmodality{k} reduces in linear time to \maxkmodality{k}.
\end{observation}

Let $\langle G, m: V(G) \rightarrow \even{4} \rangle$ be an instance of \maxmodality; also, let $\beta$ be a leaf-block of the BC-tree $\calT{}$ of $G$ and let $v$ be the parent cut-vertex of $\beta$ in $\calT{}$. We denote by $G^-_\beta$ the subgraph of $G$ induced by $v$ and the vertices of $G$ not in $\beta$, i.e., $G^-_\beta = G - (\beta - \{v\})$. Also, let $B(\calT{})$ be the set of blocks in $\calT{}$.
%
%
%
%
%
We show that \maxkmodality{k}  (and \kmodality{k}, by \cref{obs:reduction}) in simply-connected digraphs is Turing reducible to \maxkmodality{k} in biconnected digraphs. 

\begin{theorem}\label{th:simply-to-biconnected}
Given a subroutine {\sc TestBiconnected} that tests \maxkmodality{k} for \linebreak biconnected~instances, there exists a procedure  {\sc TestSimplyConnected} that tests \linebreak \maxkmodality{k} for simply-connected digraphs. Further, given an instance $\langle G, m \rangle$ of \linebreak \maxkmodality{k}, the runtime of {\sc TestSimplyConnected($\langle G, m \rangle$)} is
$$\mathcal{O}\big(|G|+ \log{k}\sum_{\beta \in B(\calT{})} r(\beta)\big),$$ where $r(\beta)$ is the runtime of  {\sc TestBiconnected($\langle \beta, m \rangle$)} and $\calT{}$ is the BC-tree of $G$. 
\end{theorem}

\begin{sketch}The algorithm selects a leaf-block $\beta$ of $\mathcal T$, with parent cut-vertex $v$, and finds an embedding of $\beta$ with the minimum modality at $v$ satisfying $m$ at all vertices, by performing a binary search using the {\sc TestBiconnected} procedure. We then remove $\beta$, replace $G$ with $G^-_\beta$, and update the value of $m(v)$, so to account for the alternations at $v$ introduced by $\beta$. The procedure terminates when all the blocks have been processed.
Therefore, since the total number of calls to subroutine {\sc TestBiconnected} is bounded by the number of blocks of $G$, which is ${O}(|\mathcal T|)={O}(|G|)$ multiplied by $\log{k}$, the overall running time is \mbox{${O}(|G|+ \log{k} \sum_{\beta \in B(\calT{})} r(\beta))$}.\end{sketch}


\subsection{Biconnected Graphs} \label{se:bico}

\ifshortversion{
To handle the decomposition of a biconnected digraph into its triconnected components, we use SPQR-trees, a data structure introduced by Di Battista and Tamassia~\cite{BattistaT90}. 

\paragraph{\bf SPQR-trees.} 
\begin{figure}[t]
\includegraphics[width=\textwidth]{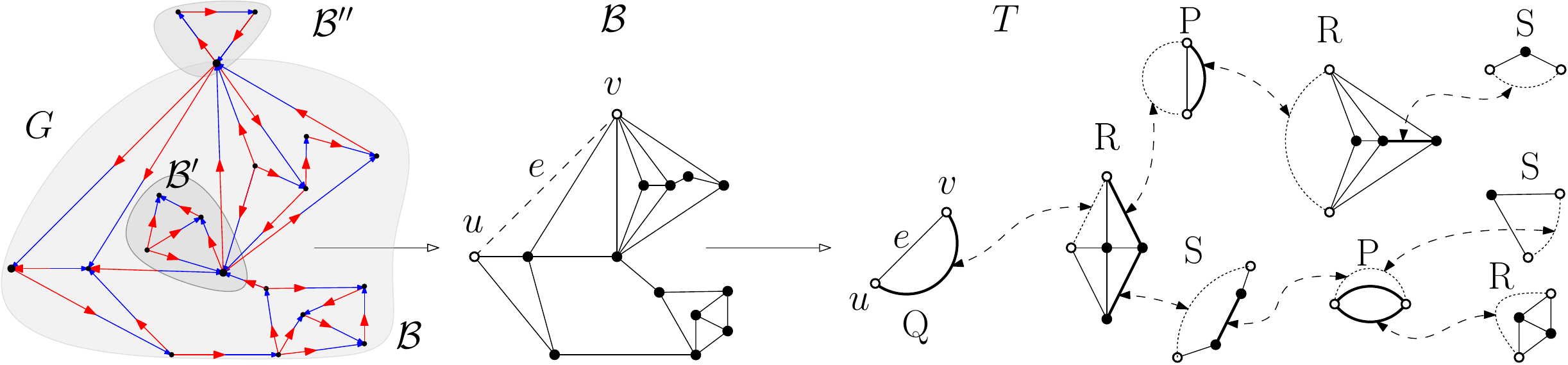}
\caption{
\ifshortversion{
(left) A $4$-modal embedding of a simply-connected planar digraph $G$. (right) The SPQR $\cal T$ of the block $\cal B$ of $G$ rooted at edge $e=uv$. The extended skeletons of all non-leaf nodes of $\cal T$ are shown; virtual edges corresponding to S-, P-, and R-nodes are thick. 
}\fi
}\label{fi:spqr}
\end{figure}
Let $G$ be a biconnected digraph. We consider SPQR-trees that are rooted at an edge $e$ of $G$, called the \emph{reference edge}. The rooted SPQR-tree $\cal T$ of $G$ with respect to $e$ describes a recursive decomposition of $G$ induced by its split pairs. The nodes of $\cal T$ are of four types: S, P, Q, and R. 
Each node $\mu$ of $\cal T$ has an associated undirected multigraph $\skel(\mu)$, called the \emph{skeleton} of $\mu$, with two special nodes $u_\mu$ and $v_\mu$ (the \emph{poles} of $\mu$), and an associated subgraph $\pert(\mu)$ of $G$, called \emph{pertinent} of $\mu$. The skeleton graph equipped with the edge $u_\mu v_\mu$, called the \emph{parent edge}, is the \emph{extended skeleton} of $\mu$. Refer to \red{ \cref{fi:spqr}(right)}. 
Each edge of $\skel(\mu)$, called \emph{virtual edge}, is associated with a child of $\mu$ in $\cal T$. The skeleton of $\mu$ describes how the pertinent graphs of the children of $\mu$ have to be ``merged'' via their poles to obtain $\pert(\mu)$. 
The extended skeleton of an S-, P-, R-, and Q-node is a cycle, parallel, triconnected graph, and a $2$-gon, respectively. It follows that skeleton and pertinent graphs are always biconnected once the parent edge is added. 
A \emph{series-parallel digraph} is a biconnected planar digraph whose SPQR-tree only contains S-, P-, and Q-nodes. A \emph{partial 2-tree} is a digraph whose every block is a series-parallel digraph.

A digraph $G$ is planar if and only if the skeleton of each R-node in the SPQR-tree of $G$ is planar. 
By selecting \emph{regular embeddings} for the skeletons of the nodes of $\cal T$, that is, embeddings in which the parent edge is incident to the outer face, we can construct any embedding of $G$ with the edge $e$ on the outer face, where the choices for the embeddings of the skeletons are all and only the (i) flips of the R-nodes and the (ii) permutations of the P-nodes.
}\fi

\smallskip
Consider a pair $\langle G, m \rangle$ such that $G$ is biconnected and let $\cal E$ be a planar embedding of $G$. Also, let $\calT{}$ be the SPQR-tree of $G$ rooted at an edge $e$ of $G$ incident to the outer face of~$\cal E$.
We will assume that the virtual edges of the skeletons of the nodes in $\calT{}$ are oriented so that the extended skeleton of each node $\mu$ is a DAG with a single source $u_\mu$ and a single sink $v_\mu$.
 \ifshortversion{}\else{This implies that the virtual edges belonging to the extended skeleton of a P-node have the same orientation, from $u_\mu$ to $v_\mu$, and that the virtual edges of the skeleton of an S-node form a directed path from $u_\mu$ to $v_\mu$.} \fi
Let $\mu$ be a node of $\calT{}$ and let $\cal E_\mu$ be the planar (regular) embedding of $\skel(\mu)$ induced by $\cal E$. 
For an oriented edge $d=uv$ of $\skel(\mu)$, the \emph{left} and \emph{right face} of $d$ in $\cal E_\mu$ is the face of $\cal E_\mu$ seen to the left and to the right of $d$, respectively, when traversing this edges from~$u$ to $v$. We define the \emph{outer left (right) face} of $\cal E_\mu$ as the left (right) face of the edge $u_\mu v_\mu$ in $\cal E_\mu$.

\paragraph{\bf Embedding tuples.}
An \emph{embedding tuple} (for short, \emph{tuple}) is a $4$-tuple $\langle \sigma_1, a, \sigma_2, b\rangle$, where $\sigma_1,\sigma_2 \in \{\uparrow, \downarrow\}$ are orientations and $a,b \in \mathds{N}$ are non-negative integers. 
Consider two tuples $t=\langle \sigma_1, a, \sigma_2, b \rangle$ and $t'=\langle \sigma'_1, a', \sigma'_2, b' \rangle$. We say that $t$ \emph{dominates} $t'$, denoted as  $t \preceq t'$, if
$\sigma_1 = \sigma'_1$, $\sigma_2 = \sigma'_2$, $a \leq a'$, and $b \leq b'$. Also, we say that $t$ and $t'$ are \emph{incompatible}, if none of them dominates the other. 
Since the relationship $\preceq$ is reflexive, antisymmetric, and~transitive,~it~defines~a~poset~$(T, \preceq)$, where $T$ is the set of embedding tuples. 
A subset $S \subseteq T$ is \emph{succinct} or an \emph{antichain}, if the tuples in $S$ are pair-wise incompatible. 
Consider two subsets $S,S' \subseteq T$ of tuples. We say that $S$ \emph{dominates} $S'$, denoted as $S \preceq S'$, if for any tuples $t' \in S'$ there exists at least one tuple $t \in S$ such that $t \preceq t'$. Also, $S$ \emph{reduces} $S'$ if  $S \preceq S'$ and $S \subseteq S'$. Finally, $S$ is a \emph{gist} of $S'$, \mbox{if $S$ is succinct and reduces~$S'$.}

Let $e_u$ and $e_v$ be the edges of $\pert(\mu)$ incident to the outer left face of $\cal E_\mu$ and to $u_\mu$ and $v_\mu$, respectively, possibly $e_u = e_v$. Also, let $a$ and $b$ be non-negative integers. 
We say that the embedding $\cal E_\mu$ \emph{realizes} tuple $\langle \sigma_1, a, \sigma_2, b\rangle$, if $\sigma_1 = \sigma(e_u, u_\mu)$, $\sigma_2 = \sigma(e_v, v_\mu)$, and $a$ and $b$ are the number of inner faces of $\cal E_\mu$ whose (two) edges incident to $u_\mu$ and to $v_\mu$, respectively, form an alternating pair. 
A tuple $t=\langle \sigma_1, a, \sigma_2, b \rangle$ is \emph{realizable by} $\mu$, if there exists an embedding of $\pert(\mu)$ that realizes~$t$, and \emph{admissible}, if $a \leq m(u)$ and $b \leq m(v)$. 
A tuple is \emph{good for} $\mu$ if it is both admissible and realizable by $\mu$.
We denote by $S(\mu)$ the gist of the set of good tuples for a node $\mu$. Let $e_\mu$ be the virtual edge representing $\mu$ in the skeleton of the parent of $\mu$ in $\calT{}$, with a small overload of notation, we also denote $S(\mu)$ by $S(e_\mu)$.
For a tuple $t=\langle \sigma_1, a, \sigma_2, b \rangle \in S(e_\mu)$, where $e=u_\mu v_\mu$, the pair $(\sigma_1, a)$ is the \emph{embedding pair} of $t$ at $u_\mu$; likewise, the pair $(\sigma_2, b)$ is the \emph{embedding pair} of $t$ at $v_\mu$.
We have the following substitution lemma.

\begin{figure}[tb!]
\centering
    \subfloat
    {\includegraphics[page=2,scale=.9]{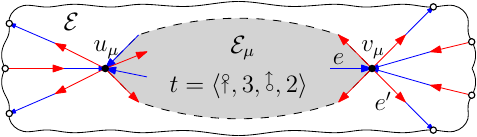}\label{fig:substitutionOrig}} \hfil
    \subfloat
    {\includegraphics[page=3,scale=.9]{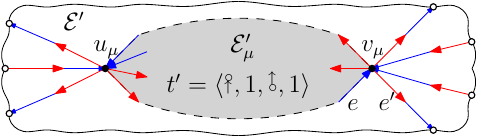}\label{fig:substitutionSwap}} 
\caption{ Illustration for the proof of \cref{le:substitution}. The parity of $t$ and $t'$ is the same at $u_\mu$ and different at $v_\mu$; in particular, even if a new alternation is introduced between the pair $(e,e')$ at~$v_\mu$, the different parity guarantees that the modality at~$v_\mu$ does not increase from $\mathcal E$ to $\mathcal E'$.
}
\label{fig:substitution}
\end{figure}

\begin{lemma}\label{le:substitution}

Let $\cal E$ be a planar embedding of $G$ satisfying $m$.
Let $\mu$ be a node of $\calT{}$ and let $\cal E_\mu$ be the embedding of $\pert(\mu)$ induced by $\cal E$. Also, let $\cal E'_\mu \neq \cal E_\mu$ be an embedding of $\pert(\mu)$ satisfying~$m$. 
Then, $G$ admits an embedding $\cal E'$ satisfying $m$ in which the embedding of $\pert(\mu)$ 
is $\cal E'_\mu$, if $t' \preceq t$, where $t$ and $t'$ are the embedding tuples realized by ${\cal E_\mu}$ and by ${\cal E'_\mu}$, respectively.
\end{lemma}

\begin{sketch}
We show how to construct a drawing $\Gamma'_G$ of $G$ satisfying $m$ in which the embedding of $\pert(\mu)$ is $\mathcal E'_\mu$; see \cref{fig:substitution}. 
Let $\Gamma_G$ be a drawing of $G$ whose embedding is $\mathcal E$. Remove from $\Gamma_G$ the drawing of all the vertices of $\pert(\mu)$ different from $u_\mu$ and $v_\mu$ and the drawing of all the edges of $\pert(\mu)$. Denote by $f$ the face of the resulting embedded graph $G^-$ that used to contain the removed vertices and edges. We obtain $\Gamma'_G$ by inserting a drawing of $\pert(\mu)$ whose embedding is $\mathcal E'_\mu$ in the interior of $f$ so that vertices $u_\mu$ and $v_\mu$ are identified with their copies in $G^-$. 
We can prove that the embedding $\mathcal E'$ of $\Gamma'_G$ satisfies $m$ by exploiting 
the interplay between the parity and the number of alternations at $u_\mu$  (at $v_\mu$) in $t'$ and $t$ when $t' \preceq t$.
\end{sketch}

\begin{fullproof}
We show how to construct a drawing $\Gamma'_G$ of $G$ satisfying $m$ in which the embedding of $\pert(\mu)$ is $\mathcal E'_\mu$; see \cref{fig:substitution}. 
Let $\Gamma_G$ be a drawing of $G$ whose embedding is $\mathcal E$. Remove from $\Gamma_G$ the drawing of all the vertices of $\pert(\mu)$ different from $u_\mu$ and $v_\mu$ and the drawing of all the edges of $\pert(\mu)$. Denote by $f$ the face of the resulting embedded graph $G^-$ that used to contain the removed vertices and edges. We obtain $\Gamma'_G$ by inserting a drawing of $\pert(\mu)$ whose embedding is $\mathcal E'_\mu$ in the interior of $f$ so that vertices $u_\mu$ and $v_\mu$ are identified with their copies in $G^-$. 

We claim that the embedding $\mathcal E'$ of $\Gamma'_G$ satisfies $m$.
First, the modality of each vertex of $G$ not in $\pert(\mu)$ is the same in $\mathcal E'$ as in $\mathcal E$. Second, the modality of each vertex in $\pert(\mu)$ different from $u_\mu$ and $v_\mu$ is the same in $\mathcal E'$ as in $\mathcal E'_\mu$. 

We only need to show that $u_\mu$ and $v_\mu$ satisfy $m$ in $\mathcal E'$.
We have that $t = \langle \sigma_1, a, \sigma_2, b\rangle $ and $t = \langle \sigma'_1, a', \sigma'_2, b'\rangle$. 
We show that the modality of $u_\mu$ in $\mathcal E'$ is smaller than or equal to the modality of $u_\mu$ in $\mathcal E$; analogous arguments hold for $v_\mu$.
We distinguish two cases. 
If $a$ and $a'$ have the same parity, as shown at vertex $u$ in \cref{fig:substitution}, then $\mathcal E'$ contains an alternating pair consisting of an edge in $G'$ and of an edge in $\pert(\mu)$ incident to $u_\mu$ only if  $\mathcal E$ contains an alternating pair consisting of an edge in $G'$ and of an edge in $\pert(\mu)$ incident to $u_\mu$. Therefore, since $a' \leq a$, the modality of $u_\mu$ in $\mathcal E'$ is smaller than or equal to the modality of $u_\mu$ in $\mathcal E$.
Otherwise, if $a$ and $a'$ have the different parity, as shown at vertex $v$ in \cref{fig:substitution}, then $\mathcal E'$ may contain an alternating pair consisting of an edge in $G'$ and of an edge in $\pert(\mu)$ incident to $u_\mu$ and to the right outer face of $\pert(\mu)$ even if $\mathcal E$ does not contain an alternating pair consisting of an edge in $G'$ and of an edge in $\pert(\mu)$ incident to $u_\mu$. 
However, in this case, it holds that  $a' < a$, therefore the modality of $u_\mu$ in $\mathcal E'$ is again smaller than or equal to the modality of $u_\mu$ in $\mathcal E$.
\end{fullproof}

Let $\calT{}$ be the SPQR-tree $\calT{}$ of $G$ rooted at a reference edge $e$.
In the remainder of the section, we show how to compute the gist $S(\mu)$ of the set of good tuples for $\mu$, for each non-root node $\mu$ of $\calT{}$. In the subsequent procedures to compute $S(\mu)$ for S-, P-, and R-nodes, we are not going to explicitly avoid set $S(\mu)$ to contain dominated tuples. In fact, this can always be done at the cost of an 
additive $O(k^2)$ factor in the running time, by maintaining an hash table that stores the tuples that have been constructed (possibly multiple times) by the procedures and by computing the gist of the constructed set as a final step.

\begin{property}\label{obs:size}
For each node $\mu \in \calT{}$, it holds that $|S(\mu)| \in O(k)$.
\end{property}

\begin{proof}
By the definition of gist, any embedding pair $(\sigma,a)$ has at most two tuples $t',t'' \in S(\mu)$ such that $(\sigma,a)$ is the embedding pair of $t'$ and $t''$ at $u_\mu$; also, the embedding pairs $(\sigma',a')$ of $t'$ and $(\sigma'',a'')$ of $t''$ at $v_\mu$ are such that $\sigma' \neq \sigma''$. Since there exist at most $2k$ realizable embedding pairs $(\sigma,a)$ at $u_\mu$ (as $\sigma \in \{\inn,\out\}$, $a \in \{0,1,\dots,k\}$, and the existence of tuple whose embedding pair at $u_\mu$ is $(\sigma,0)$ implies that all tuples have such an embedding pair at $u_\mu$), we have $|S(\mu)| \leq 4k$.\end{proof}

If $\mu$ is a leaf Q-node in $\calT{}$, then $S(\mu)=\{ \langle \sigma(u_\mu v_\mu), 0,  \sigma(u_\mu v_\mu), 0 \rangle \}$.
If $\mu$ is an internal node of~$\calT{}$, we visit $\calT{}$ bottom-up and compute the set $S(\mu)$ for $\mu$ assuming to have already computed the sets $S(\mu_1),\dots, S(\mu_k)$ for the children $\mu_1,\dots,\mu_k$ of $\mu$ 
(where $\mu_i$ is the child of $\mu$ corresponding to the edge $e_i$ in $\skel(\mu)$).
Let $\rho$ be the unique child of the root of $\calT{}$.
Once the set $S(\rho)$ has been determined, we can efficiently decide whether $G$ admits an embedding satisfying $m$ in which the reference edge $e$ is incident to the outer face by means of the following lemma.

\begin{lemma}\label{le:rho}
Given $S(\rho)$, we can test whether $G$ has an embedding that satisfies $m$ in $O(k^2)$ time. 
\end{lemma}

\begin{fullproof}
Let $t=\langle \sigma_1,a,\sigma_2,b \rangle$ be a realizable tuple in $S(\rho)$ and let ${\cal E}_t$ be the embedding of $G$ obtained by inserting the reference edge $e$ in the outer face of a regular embedding $\cal E_\rho$ of $\pert(\rho)$ realizing $t$ that satisfies $m$. Let $u$ and $v$ be the poles of $\rho$. We have the following claim.

\begin{claim}\label{cl:e_t}
Embedding ${\cal E}_t$ satisfies $m$ if and only if:
\begin{enumerate}[(i)]
    \item
\begin{inparaenum}
    \item $m(u)\geq a+1$, if $a$ is odd, or
    \item $m(u)\geq a$, if $a$ is even and $\sigma_1=\sigma(e)$, \linebreak or
    \item $m(u)\geq a+2$, if $a$ is even and $\sigma_1 \neq \sigma(e)$,
\end{inparaenum} 
and
\item
\begin{inparaenum}
    \item $m(v)\geq b+1$, if $b$ is odd, or
    \item $m(v)\geq b$, if $b$ is even and $\sigma_2=\sigma(e)$, \linebreak or
    \item $m(v)\geq b+2$, if $b$ is even and $\sigma_2 \neq \sigma(e)$.
\end{inparaenum} 
\end{enumerate}
\end{claim}
\begin{fullproof}
\begin{figure}[tb!]
\centering
\subfloat[]{\includegraphics[page=1,height=.24\textwidth]{rootNode}}\hfil
\subfloat[]{\includegraphics[page=2,height=.24\textwidth]{rootNode}}\hfil
\subfloat[]{\includegraphics[page=3,height=.24\textwidth]{rootNode}}
\caption{Illustrations for the proof of \cref{cl:e_t}. (a) The root edge $e$ and $\rho$ (where the orientation of $\rho$ being arbitrary). (b) A $4$-modal embedding of $G$, where $u$ has $2$ alternations and $v$ has $4$ alternations. (b) A $6$-modal embedding of $G$, where $u$ has $4$ alternations and $v$ has $6$ alternations.}
\label{fi:rootNode}
\end{figure}
To prove the statement for vertex $u$, we just need to observe that, if $a$ is odd, then the edges $e'_u$ and $e''_u$ incident to $u$ and to the left and to the right outer face of ${\cal E}_\rho$, respectively, have opposite orientations, while if $a$ if even, then these edges have the same orientation (in this case, possibly $e'_u = e''_u$). Also, since $a$ is the number of alternations between edges incident to $u$ and to the internal faces of ${\cal E}_\rho$, the modality at $u$ in ${\cal E}_\rho$ is equal to $a+1$, if $a$ is odd, while it is equal to $a$, otherwise.
Therefore, since edge $e$ appears between $e'_u$ and $e''_u$ in embedding ${\cal E}_t$ obtained from ${\cal E}_\rho$, we have that the number of alternations around $u$ in ${\cal E}_t$ is the same as in ${\cal E}_\rho$, if $a$ is odd or if $a$ is even and $\sigma(e) = \sigma_1$, and it is equal to the modality of $u$ in ${\cal E}_\rho$ plus $2$, if $a$ is even and $\sigma(e) \neq \sigma_1$. Refer to  \cref{fi:rootNode}. The proof of the statement for vertex $v$ is analogous.
\end{fullproof}

By  \cref{cl:e_t}, for each realizable tuple $t \in S(\rho)$, we can test whether embedding  ${\cal E}_t$ satisfies $m$ in constant time. Also, $|S(\rho)| \in O(k^2)$, by \cref{obs:size}. Therefore, we can test in $O(k^2)$ time whether there exists an embedding of $\pert(\rho)$ that can be extended to an embedding of $G$ that satisfies $m$ in which $e$ is incident to the outer face.
\end{fullproof}

\ifshortversion{}\else{We remark that the choice of the reference edge does not affect the existence of a $4$-modal embedding. In fact, a change of the outer face such that edge $e$ is incident to such a face can always be performed while preserving the rotation at any vertex.}\fi

\section{Partial 2-trees} \label{se:2trees}
In the following, we describe how to compute $S(\mu)$, if $\mu$ is an S-node (\cref{le:S}) and a P-node (\cref{le:P}) in ${O}(f(k)|\skel(\mu)|)$ time, where $f$ is a computable function.

\begin{lemma}\label{le:S}
Set $S(\mu)$ can be constructed in ${O}(k^2|\skel(\mu)|)$ time for an S-node $\mu$.
\end{lemma}

\begin{sketch}
Let $\mu$ be an S-node with skeleton $\skel(\mu)=(e_1,e_2,\dots,e_h)$. We define $\tau_j$ as the S-node obtained by the series composition of $\mu_1, \mu_2,\dots,\mu_j$, with $j \leq h$. Initially we set $S(\tau_1)=S(e_1)$. Then, we construct $S(\tau_j)$ via dynamic programming, for $j=2,\dots,h$, by 
verifying the compatibility of the embedding pairs of the good tuples of the virtual edges of $\skel(\tau_j)$ at the internal vertices of $\skel(\tau_j)$. As $S(\tau_j)=S(\tau_{j-1}) \cup e_j$, we can compute $S(\tau_j)$ by considering all the tuples obtained by combining 
every tuple $t' \in S(\tau_{j-1})$ with every tuple $t'' \in S(e_j)$. Since both these sets contain $O(k)$ tuples, by \cref{obs:size}, and since the tuple resulting from the combination of $t'$ and $t''$ can be determined in $O(1)$ time, we have that $S(\tau_j)$ can be computed in $O(k^2)$ time. Therefore, the overall running \mbox{time for computing $S(\mu)=S(\tau_h)$ is ${O}(k^2 |\skel(\mu)|)$.}
\end{sketch}

\begin{lemma}\label{le:P}
Set $S(\mu)$ can be constructed in $O((2k+4)!k^3 + |\skel(\mu)|)$ time for a P-node~$\mu$.
\end{lemma}

\begin{sketch}
Let $\mu$ be a P-node with poles $u_\mu$ and $v_\mu$, whose skeleton $\skel(\mu)$ consists of $h$ parallel virtual edges $e_1,e_2,\dots,e_h$. 
It can be shown that the computation of $S(\mu)$ reduces in $O(|\skel(\mu)|)$ time to the computation of $S(\tau)$, where $\tau$ is a P-node whose skeleton consists of at most $2k$ virtual edges of $\skel(\mu)$ that contribute with at least one alternating pair of edges at~$u_\mu$ or~$v_\mu$, plus up to $4$ virtual edges of $\skel(\mu)$ that contribute with no alternating pair at~$u_\mu$ or at~$v_\mu$. 
For any permutation $\pi$ of the virtual edges of $\pert(\tau)$, let $\tau^\pi_i$ be the P-node obtained by restricting $\tau$ to the first~$i$ virtual edges in $\pi$. 
We fix the embedding of $\skel(\tau^\pi_i)$ in such a way that the virtual edges of $\skel(\tau^\pi_i)$ are ordered according to $\pi$.
Then, in a fashion similar to the S-node case, we can compute~$S(\tau^\pi_i)$ for the given embedding of $\skel(\tau^\pi_i)$ by combining $S(\tau^\pi_{i-1})$ and $S(e_i)$ in $O(k^2)$ time (recall that both these sets have size $O(k)$, by \cref{obs:size}). 
Clearly, for any fixed $\pi$, we can compute $S(\tau^h_\pi)$ in $O(k^3)$ time. Thus, by performing the above computation for all the $(2k+4)!$ possible permutations for the virtual edges of $\pert(\tau)$, we can construct $S(\tau)$ in $O((2k+4)!k^3+|\skel(\mu)|)$ time.\end{sketch}

Altogether, \cref{le:S,le:P} yield the following main result.

\begin{theorem}\label{th:sp}
\maxkmodality{k} can be solved in $O((2k+4)!k^3  n)$ for series-parallel digraphs.
\end{theorem}

 \cref{obs:reduction}, \autoref{th:simply-to-biconnected}, and \autoref{th:sp} immediately imply the following.

\begin{corollary}\label{co:2trees}
\kmodality{k} can be solved in $O(((2k+4)!k^3\log{k})  n)$ for directed partial $2$-trees.
\end{corollary}

Due to the special algorithmic framework we are employing,
 we can however turn the multiplicative $O(\log{k})$ factor in the running time into an additive $O(k)$ factor by modifying \ifshortversion{\cref{th:simply-to-biconnected} }\else{\cref{algo:simply-to-bico} }\fi as follows. When considering a cut-vertex $v$\ifshortversion{, }\else{ (line~$4$), }\fi we will execute ``only once'' the function {\sc TestBiconnected} \ifshortversion{}\else{(line~$5$) }\fi 
 by rooting the SPQ-tree at a Q-node $\eta$ corresponding to an edge incident to~$v$. This will allow us to compute the minimum modality for cut-vertex $v$ in an embedding that satisfies $m$ at every vertex, by simply scanning the set $S(\eta)$, which takes $O(k)$ time by \cref{obs:size}, rather than by exploiting a logarithmic number of calls to {\sc TestBiconnected}. 

\begin{theorem}\label{th:2trees-efficient}
\kmodality{k} can be solved in $O((2k+4)!k^3 n)$ for directed partial $2$-trees.
\end{theorem}


\section{A Linear-time Algorithm for 4-MaxModality when $\mathbf{\Delta \leq 6}$} \label{se:deg6} 

In this section, we show that in the special case when $k=4$ and $G$ has maximum degree~$\Delta \leq 6$, it is possible to compute the set $S(\mu)$ when $\mu$ is an R-node in linear time in the size of $\skel(\mu)$. 
Our strategy to compute $S(\mu)$ is as follows.  We select a single tuple from the admissible set of each virtual edge incident to $u_\mu$ and $v_\mu$, in every possible way. Each selection determines a ``\emph{candidate tuple}'' $t$ for $S(\mu)$. 
First, we check if $t$ is admissible at both $u$ and $v$. 
Second, we restrict the tuples of the edges incident to the poles to only the tuples that form $t$ and check if there is a way of satisfying $m$ at the (inner) vertices of $\skel(\mu)$.
If both the poles and the inner vertices are satisfiable, then we add $t$ to $S(\mu)$. Since the degrees of the poles are bounded, there is at most a constant number of candidate tuples which must be checked. The complexity lies in this check.

    We now formally describe how to compute $S(\mu)$.
First, for each virtual edge $e_i$ of $\skel(\mu)$ incident to the poles of $\mu$, we select a tuple $t_i$ from $S(\mu_i)$. 
Let $T_u = [t_{u,1},t_{u,2},\dots,t_{u,\ell}]$ and $T_v=[t_{v,1},t_{v,2},\dots,t_{v,h}]$ be the list of tuples selected for the virtual edges incident to $u_\mu$ and to $v_\mu$, respectively.
Each pair of lists $T_u$ and $T_v$ yields a {\em candidate tuple} $t = \langle \sigma_1, a, \sigma_2, b \rangle$ for $\mu$.
However, the tuples selected to construct $T_u$ and $T_v$ allow for an admissible embedding of $\pert(\mu)$ \mbox{realizing tuple $t$ {\em if and only if}:}
\begin{inparaenum}
    \item[{\bf (Condition~1)}] tuple $t$ satisfies $m$ at $u_\mu$ and at $v_\mu$, and
    \item[{\bf (Condition~2)}] it is possible to select tuples for each of the remaining virtual edges of $\skel(\mu)$ that satisfy $m$ at every internal vertex of $\skel(\mu)$.
\end{inparaenum}
Let $\mathcal P(\mu)$ be the set of candidate tuples for $\mu$ constructed as described above.
We can easily filter out the candidate tuples that do not satisfy Condition~1.
For each pair of lists $T_u$ and $T_v$ yielding a tuple $t \in \mathcal P(\mu)$, we will show how to test Condition~2 for $\mu$ in linear time. This and the fact that $|\mathcal P(\mu)| \in O(1)$ imply the following.

\begin{lemma}\label{le:R}
Set $S(\mu)$ can be constructed in ${O}(|\skel(\mu)|)$ time for an \mbox{R-node}~$\mu$, if $\Delta\leq 6$.
\end{lemma}

Altogether, \cref{le:S,,le:P,,le:R} yield the following main result.

\begin{lemma}\label{le:bico-max-degree}
\maxkmodality{4} can be solved in linear time for biconnected digraphs with $\Delta \leq 6$.
\end{lemma}

\cref{obs:reduction}, \cref{th:simply-to-biconnected}, and \cref{le:bico-max-degree} immediately imply the following.

\begin{theorem}\label{th:max-degree}
\modality{} can be solved linear time for digraphs with $\Delta \leq 6$.
\end{theorem}

To prove \cref{le:R}, we show how to solve the following auxiliary problem for special instances.

\problemdef{\sc $4$-MaxSkelModality}{A triple $\langle G=(V,E), {\cal S}=\{S(e_1),\dots,S(e_{|E|})\}, m \rangle$ where
$G$ is an embedded directed graph, each $S(e_i)$ is a set containing embedding tuples for the virtual edge $e_i \in E$, and $m: V \rightarrow \even{4}$ is the \emph{maximum-modality function}.}{
Can we select a tuple from each set $S(e_i)$ in such a way that the modality at each vertex $v \in V$ is at most $m(v)$?}

\medskip
For each pair of lists $T_u$ and $T_v$ yielding a candidate tuple in $\mathcal P(\mu)$, we will construct an instance $I_\mu(T_u,T_v)=(G,\mathcal S, m)$ of {\sc $4$-MaxSkelModality} as follows.
\begin{inparaenum}
\item We set $G=\skel(\mu)$ and we fix the embedding of $G$ to be equal to the unique regular embedding of $\skel(\mu)$;
\item for each virtual edge $e_{u,i}$ incident to $u_\mu$, with $i=1,\dots,\ell$, we set $S(e_{u,i})=\{t_{u,i}\}$; for each virtual edge $e_{v,j}$ incident to $v_\mu$, with $j=1,\dots,h$, we set $S(e_{v,j})=\{t_{v,j}\}$; for each of the remaining virtual edges $e_d$ of $\skel(\mu)$, we set $S(e_d) = S(\mu_d)$; finally,
\item \mbox{the maximum-modality function of $I_\mu$ coincides with $m$.}
\end{inparaenum}

Clearly, $I_\mu(T_u,T_v)$ is a positive instance of {\sc $4$-MaxSkelModality} if and only if, given the constrains imposed by the tuples in $T_u$ and in $T_v$, there exists a selection of tuples for the edges of $G$ not incident to $u_\mu$ or $v_\mu$ that satisfies $m$ at all the \mbox{internal vertices of $G$, i.e., Condition~2 holds.}

Let $v$ be a vertex of $G$ and let $e$ be an edge in $E(v)$, we denote by $A_v(e)$ the maximum number of alternations at $v$ over all the tuples in~$S(e)$.

\begin{definition}[Good instances]\label{prop:max6}
An instance of {\sc $4$-MaxSkelModality} is \emph{good} if, for any vertex $v$ in $G$, it holds 
$\sum_{e\in E(v)} (A_v(e)+1) \leq 6$.
\end{definition}

Note that, for each edge $e$ in $ske(\mu)$ incident to a vertex $v$, $\pert(e)$ contributes at least $A_v(e)+1$ edges to $d_{\pert(e)}(v)$. Thus, we have
$\sum_{e\in E(v)} (A_v(e)+1) \leq \sum_{e\in E(v)} d_{\pert(e)}(v) \leq 6$. 
Therefore, instance $I_\mu(T_u,T_v)$ is good.
Although {\sc $4$-MaxSkelModality} turns out to be NP-complete in general \ifshortversion{(\appendixHardness)}\else{(\cref{th:npc})}\fi, we are now going to show the following main positive result.

\begin{theorem}\label{th:ptime-rnode}
{\sc $4$-MaxSkelModality} is linear-time solvable for good instances.
\end{theorem}

The outline of the linear-time algorithm to decide whether a good instance $I=\langle G=(V,E), {\cal S}=\{S(e_1),\dots,S(e_{|E|})\}, m \rangle$ of {\sc $4$-MaxSkelModality} is a positive instance is a follows\ifshortversion{.}\else{; see also \cref{algo:reduceGraph}.}\fi

\begin{itemize}[-]
    \item We process $I$ by means of a set of \emph{reduction rules} applied locally at the vertices of $G$ and their incident edges. Each of these rules, if applicable, either detects that the instance $I$ is a negative instance or transforms it into an equivalent smaller instance~$I' = \langle G', {\cal S}', m'\rangle$. 
    Each rule can be applied when specific conditions are satisfied at the considered vertex. A rule may additionally set a vertex as \emph{marked}. 
    Any marked vertex $v$ has the main property that {\bf any selection} of tuples from the admissible sets of the edges \mbox{incident to~$v$ satisfies~$m'$ at~$v$.} 
    \item  Let $I^*$ be the instance of {\sc $4$-MaxSkelModality} obtained when no reduction rule may be further applied.
    We prove that instance $I^*$ has a special structure that allows us to reduce the problem of testing whether $I^*$ is a positive instance of {\sc $4$-MaxSkelModality} to that of verifying the \NAE-satisfiability of a constrained instance of {\sc NAESAT}, in fact, of {\sc Planar NAESAT}.
    Since {\sc Planar NAESAT} is in P~\cite{Moret1988}, this immediately implies that {\sc $4$-MaxSkelModality} is also in P. 
    However, in \ifshortversion{\appendixNAESAT}\else{ \cref{sse:naesat}}\fi, 
    by strengthening a result of Porschen \etal~\cite{PorschenRS03}, we are able to show that the constructed instances of {\sc NAESAT} are always satisfiable and that a satisfying \NAE-truth assignment can \mbox{be computed in linear~time.}
\end{itemize}

In \appendixIrreducible{}, we provide three reduction rules that turn a good instance $I$ into an equivalent smaller good instance $I'$. 
Let $I^* = \langle G^*, \mathcal S^*, m^* \rangle$ be the good instance, equivalent to $I$, produced by applying a maximal sequence of reduction rules to~$I$.
We say that $I^*$ is \emph{irreducible}.

The following lemma will prove useful.

\begin{lemma}\label{le:unmarked}
For each unmarked vertex $v \in V(G^*)$, it holds that:
\begin{inparaenum}[(i)]
\item $v$ has degree 3,
\item $m^*(v)=4$, and
\item there exist tuples $t_1,t_2 \in S^*(e)$ such that the embedding pair of $t_1$ and of $t_2$ at $v$ are $(\inn, 1)$ and $(\out, 1)$, respectively, for each edge $e$ incident to $v$.
\end{inparaenum}
\end{lemma}

\begin{fullproof}
Let $v$ be an unmarked vertex of $G^*$. Since $I^*$ is irreducible, none of the conditions of Rules~1,~2, and~3 apply at $v$.
We denote by $A^*_x(e)$ the maximum number of alternations of the embedding tuples of $S^*(e)$, where $x$ is a vertex incident to the edge $e$.

We first show that $|E^*(v)| = 3$ by contradiction.
Suppose first that $A^*_v(e)=0$ for all the edges incident in $E^*(v)$, then the variety of these edges at $v$ is $1$ and there is only one combination of tuples at $v$. Therefore, either {\sc Rule}~1 would have rejected the instance, or {\sc Rule}~2 would have marked $v$, contradicting to the fact that $I^*$ is irreducible. 
Suppose now that $E^*(v)$ contains exactly one edge $e$ with $A^*_v(e)>0$, then each tuple of $S^*(e)$ participates in a single combination. It follows that either one of the combinations is unsatisfying and {\sc Rule}~1 would have applied, or they are all satisfying and {\sc Rule}~2 would have marked $v$, contradicting to the fact that $I^*$ is irreducible. 
Finally, suppose that $E^*(v)$ contains exactly two edges $e_1$ and $e_2$ with $A^*_v(e_1)>0$ and $A^*_v(e_2) > 0$, then the rest of the edges of $E^*(v)$ have variety $1$ at $v$ so {\sc Rule}~3 would have applied. As $I^*$ is irreducible, however, this yields a contradiction.
Therefore, $E^*(v)$ must contains at least $3$ edges $e_1,e_2,e_3$ with $A^*_v(e_1),A^*_v(e_2),A^*_v(e_3) \geq 1$. Furthermore, since $I^*$ is a good instance $\sum_{e\in E^*(v)} (A^*_v(e)+1) \leq 6$, so there can be at most $3$ such edges, proving Property~(i) of the statement.

The arguments above also imply that $A^*_v(e_1) = A^*_v(e_2) = A^*_v(e_3) = 1$. This, there must be at least $3$ alternations at $v$, and since $v$ is still unmarked and the maximum-modality is even, $m^*(v)$ must be $4$, proving Property~(ii) of the statement.

Finally, we prove Property (iii) of the statement.
First, using the same arguments used to prove Property (i), we have that if all or all but one of these edges have variety $1$ at $v$ then {\sc Rule}~1 or {\sc Rule}~2 apply. Likewise, if exactly one of these edge has variety $1$ at $v$, then the other two edges have variety greater than $1$ at $v$ and {\sc Rule}~3 applies.
Thus, all three edges must have a variety of at least $2$ at $v$. In fact, as they contribute with exactly one alternation at $v$, by Property (ii), they must have variety exactly $2$. Since $(\inn, 1)$ and $(\out, 1)$ are the only two embedding pairs with a single alternation, this proves Property (iii). 
\end{fullproof}

Our next and final tool is the following, quite surprising, result.

\begin{lemma}\label{lem:irreducible}
Any irreducible good instance $I^*$ is a positive instance. 
\end{lemma}

\autoref{th:ptime-rnode} immediately follows from \cref{lem:irreducible}.\ifshortversion{
We conclude the section by providing a sketch of the proof of \cref{lem:irreducible}. A detailed proof can be found in \appendixIrreducible.

\paragraph{Outline of the proof of \cref{lem:irreducible}.} 
If a vertex is marked then any combination of tuples will satisfy $m^*$ at it. So the proof is mainly concerned with unmarked vertices. 
By \cref{le:unmarked}, edges where both endpoints are unmarked have one of the following tuple sets: $S_A = \{\langle \downarrow, 1, \downarrow, 1 \rangle, \langle \uparrow, 1, \uparrow, 1 \rangle\}$, $S_B = \{\langle \uparrow, 1, \downarrow, 1 \rangle, \langle \downarrow, 1, \uparrow, 1 \rangle \}$, or $S_A \cup S_B$. In the last case we arbitrarily remove either $S_A$ or $S_B$.
Taking advantage of the structure of irreducible instances, the problem of solving $I^*$ is reduced in linear time to the one of testing the \NAE-satisfiability of a CNF-formula $\phi$ in which every variable occurs in at most two clauses. 
Each edge incident to an unmarked vertex has the two possible embedding pairs $(\inn,1)$ or $(\out,1)$ at the vertex. We create a variable for each incidence between an edge and an unmarked vertex.
For each edge with two unmarked endpoints, we introduce an \emph{edge clause} to ensure that the embedding pairs for each endpoint are selected in a consistent way.
Consider an unmarked vertex $v$ and assume, for simplicity of description, that its three incident edges have the same orientation at $v$.
A selection of embedding pairs for the edges incident to $v$ will not satisfy $m^*(v)$ if and only if all such pairs coincide. Therefore, we can introduce a \emph{vertex clause} to model such constraint as a \NAESAT clause that is the disjunction of the three boolean variables for the endpoints of the edges incident to $v$.
The \NAE-formula $\phi$ has the property that each variable occurs in at most two clauses. Moreover, the variable-clause graph $G_\phi$ of $\phi$ contains no connected component that is isomorphic to a simple cycle, since vertex clauses have degree $3$.
 In \appendixNAESAT, we prove that such instances are always \NAE-satisfiable and provide a linear-time algorithm to construct a \NAE-truth assignment for such formulas. \mbox{This proves that $I^*$ is always a positive instance.}
}
\else
{ \cref{se:irreducible} is devoted to the proof of \cref{lem:irreducible}.
}
\fi


\section{Conclusions}\label{se:conclusions}

In this paper, we studied the complexity of the \kmodality{k} problem, with special emphasis on $k=4$. We provided complexity, algorithmic, and combinatorial results. 
Our main algorithmic contribution for $k=4$ and $\Delta \leq 6$ leverages an elegant connection with the \NAE-satisfiability of special CNF formulas, whose study allowed us to strengthen a result in~\cite{PorschenRS03}.
Moreover, we showed notable applications of the previous results to some new interesting embedding problems for clustered networks, some of which solve open problems \mbox{in this area~\cite{DBLP:journals/jgaa/AngeliniLBFPR17,GiacomoLPT17}}. 

\bibliographystyle{abbrv}  
\bibliography{bibliography}
\end{document}